# Global alignment for relation extraction in Microbiology

Anfu TANG, Claire Nédellec, Pierre Zweigenbaum, Louise Deléger, Robert Bossy

MaIAGE, INRAE, Université Paris-Saclay, 78350 Jouy-en-Josas, France
Limsi, CNRS, Université Paris-Saclay, 91400 Orsay, France
`anfu.tang@inrae.fr`

**Abstract.** We investigate a method to extract relations from texts based on global alignment and syntactic information. Combined with SVM, this method is shown to have a performance comparable or even better than LSTM on two RE tasks.

**Keywords:** Natural Language Processing (NLP), Relation extraction (RE), Shortest dependency path, Global Alignment (GA), Support Vector Machine (SVM), Long Short-Term Memory (LSTM)

## 1   Motivation

In microbiology, there exists a large amount of documents in which relations between different concepts are the most important information for scientists. Accessing this information demands expert knowledge and time. Relation extraction addresses this issue by proposing automated methods to detect relations of interest and thus plays an important role in the field of Natural Language Processing (NLP). The aim of relation extraction is to categorize the relations between certain pairs of named entities, which are terms of importance pre-tagged according to the application domain. In this article, entities in the microbiological domain are targeted. For example, in the sentence "The persistence of the fish pathogen [*Vibrio salmonicida*]$_{e1}$ in [*fish farm sediments*]$_{e2}$ was studied by use of fluorescent-antibody techniques", the entities *Vibrio salmonicida* and *fish farm sediments* are linked by the relation Lives_In (e1,e2).

Traditional machine-learning methods for relation extraction are usually dependent on kernel design or feature engineering, while neural methods have exhibited greater performance in related NLP tasks in recent years. However, a drawback of neural methods is that they require large training corpora. In specialized domains such as microbiology, training corpora are often of limited size, and neural methods may not generalize well. Thus, it is worth investigating less demanding classic methods as an alternative.

## 2   Related work

Syntactic information, represented as dependencies between words, is particularly relevant for relation extraction. More specifically, the shortest dependency path (SDP) between two entities is considered to contain the most relevant information of the relation. Given a dependency tree, we can obtain the SDP through Dijkstra's algorithm.

To reduce data sparsity, words along a SDP are usually represented by word features (POS tags, lemmas, etc). Bunescu and Mooney (2005) proposed a kernel by which the similarity between two SDPs is represented by the product of the number of shared word features at each position. However, this algorithm produces a null similarity when two paths have different lengths. To address this issue, Valsamou (2014) built upon a method proposed by Philippe Veber (philippe.veber@univ-lyon1.fr) to align SDP pairs, using the Global Alignment algorithm of Needleman and Wunsch (1970). More recently Xu et al. (2015) used SDP with Long Short-Term Memory network (LSTM).

## 3 Proposed method

Based on the hypothesis that classic machine-learning methods may outperform deep-learning methods on a small dataset, we propose a method based upon Support Vector Machines. We started from Valsamou's method and removed its dependencies to manually designed, domain-specific information and to a specific parser. The method uses an Empirical Kernel Map (Schölkopf and Smola, 2002), i.e., a similarity matrix between pairs of instances. Its contribution concerns the similarity function used to compute this matrix, SDP-GA. SDP-GA leverages the shortest dependency path (SDP) between two entities and computes Global Alignment (GA) to compare two SDPs.

We reimplemented this method with several modifications and applied it to two RE datasets. In more detail, sentences are first parsed into dependency trees, then the SDPs are extracted as input data for each pair of entities. Empirical Kernel Map is used to transform SDPs into numeric features. There are two steps: 1) each SDP is encoded by a vector containing its similarity with each SDP in the training set, computed using the Global Alignment algorithm; 2) A standard kernel is applied to obtain a kernel matrix.

We compare the SVM method to a neural method, SDP-LSTM (Xu et al., 2015). Its code was not available, we reimplemented and updated it. The SDP-LSTM model leverages the SDP with long short-term memory (LSTM) units. Each SDP is represented at four levels: word sequence, represented as pre-trained fastText word embeddings, POS tag sequence, dependency tag sequence, WordNet hypernym sequence, each represented as trainable embeddings. These sequences are fed into four channels. For each channel, a small LSTM network is built, then the outputs are concatenated and used for final prediction.

## 4 Experiments

We focus on two tasks. The first task is the BB-rel task of the Bacteria Biotope Task at BioNLP Open Shared Tasks 2019 (Bossy et al., 2019). There are two relation types: Relation Lives_In between Microorganism and Habitat/Geographical entities; Relation Exhibits between Microorganism and Phenotype entities. The second task is the LLL challenge (Nédellec, 2005) which aims to extract pro-tein/gene interactions from biological abstracts. There is only one relation type.

Table 1 summarizes the F1-scores obtained by the two methods on the two datasets.

Table 1. Comparison of F1-scores (%) obtained on the two shared task datasets.

| Method/Dataset | BB-rel 19 | LLL |
|---|---|---|
| SDP-GA-SVM | 61.7 | **64.4** |
| SDP-LSTM | 65.0 | 50.2 |
| Best BB-rel 19 participant [Xiong et al., 2019] | **66.4** | N/A |
| Best LLL 2005 participant [Riedel and Klein, 2005] | N/A | 54.3 |

## 5   Discussion and Conclusion

Results show that the SVM method obtains much better performance (+14.2pt F1) on the smaller corpus (LLL, 926 training examples) whereas the LSTM method obtains better performance (+3.3pt F1) on the larger corpus (BB-rel, 2434 training examples). This confirms our hypothesis that on small corpora, it is worth studying further kernel-based methods coupled with the SDP-GA. As future work, we plan to explore more recent transfer-learning based neural methods such as BERT-style Transformers.